\documentclass[a4paper, 10pt, conference]{ieeeconf}       
\newtheorem{thm}{Theorem}[section]
\newtheorem{defn}[thm]{Definition}

\newtheorem{rem}[thm]{Remark}
\newtheorem{prop}[thm]{Proposition}

\IEEEoverridecommandlockouts                              
\usepackage{amsfonts}
\usepackage{amsmath}
\usepackage{graphicx}      
\usepackage{float} 
\usepackage{psfrag} 
\usepackage{url}
\usepackage{subfig}
\usepackage{cite}

\title{\LARGE \bf  Sparse approximate inverses of Gramians and impulse response matrices of large-scale interconnected systems}

\author{Aleksandar Haber and Michel Verhaegen
\thanks{A. Haber and M. Verhaegen are with Delft Center for Systems and Control, Delft University of Technology, Delft, 2628 CD, The Netherlands, (e-mail: {\tt\small a.haber@tudelft.nl}; {\tt\small m.verhaegen@tudelft.nl}).}
}
\begin{document}
\maketitle
\thispagestyle{empty}
\pagestyle{empty}
\maketitle
\begin{abstract}
In this paper we show that inverses of well-conditioned, finite-time Gramians and impulse response matrices of large-scale interconnected systems described by sparse state-space models, can be approximated by sparse matrices. The approximation methodology established in this paper opens the door to the development of novel methods for distributed estimation, identification and control of large-scale interconnected systems. The novel estimators (controllers) compute local estimates (control actions) simply as linear combinations of inputs and outputs (states) of local subsystems. The size of these local data sets essentially depends on the condition number of the finite-time observability (controllability) Gramian. Furthermore, the developed theory shows that the sparsity patterns of the system matrices of the distributed estimators (controllers) are primarily determined by the sparsity patterns of state-space matrices of large-scale systems. The computational and memory complexity of the approximation algorithms are $O(N)$, where $N$ is the number of local subsystems of the interconnected system. Consequently, the proposed approximation methodology is computationally feasible for interconnected systems with an extremely large number of local subsystems.
\end{abstract}
\section{Introduction}
A large variety of estimation and control algorithms are based on the inversion of the finite-time observability (controllability) Gramians and impulse response matrices. For example, the norm-optimal Iterative Learning Control (ILC) algorithms \cite{Bristow:2006,son2013robust,pipeleers2014,Haber:13} determine a control action by computing a regularized pseudo-inverse of the impulse response matrix. The Moving Horizon Estimation (MHE) methods \cite{alessandri2003,Haber:13mhe,farina2010automatica} compute a state estimate by inverting the finite-time observability Gramian.  
In this paper the finite-time observability (controllability) Gramians and impulse response matrices are called the \textit{fundamental matrices of dynamical systems}.
\\
Large-scale interconnected systems (dynamical networks) consist of a large number of local dynamical subsystems that are interconnected in a spatial domain \cite{massioni2009,pakazad2014distributed,matni2014low,khan2008distributing,motee2008,justin2010distributed,dandrea2003,keviczky2006decentralized,bamieh02distributedcontrol,yo0,Pasqualetti2014,summers2013optimal}. In this paper, we focus on large-scale systems with interconnection structures described by sparse graphs. That is, we focus on large-scale systems described by sparse state-space models (state-space models with sparse matrices). A large number of interconnected systems are described by sparse state-space models. Some notable examples are systems obtained by discretizing Partial Differential Equations (PDEs) using the finite difference or finite element methods \cite{haberThesis,benner2004,bikcora2013acc,haber2013OL}, power systems \cite{kundur1994power,vedran2013,broussolle} and deformable mirrors for extremely large telescopes \cite{massioni2011fast,fraanje2010decomposition}. Fundamental matrices of these systems are also sparse. \textit{However, in the general case, inverses of fundamental matrices are completely dense\cite{chow2000priori}. In other words, inversion "destroys" the structure of a large-scale system.} From the estimation point of view, this means that the state of a local subsystem is a linear combination of outputs and inputs of all local subsystems in the network. That is, to compute its local state, a local subsystem needs to receive and process input-output data of all local subsystems in the network. Due to a large-scale nature of dynamical networks and because of various computational and communication constraints that are present in practice, it is often impossible to compute a local state as a linear combination of all local inputs and outputs of a dynamical network. Moreover, in many cases it might be impossible to explicitly compute the inverses of the fundamental matrices. The main reason is that the computational and memory complexity of matrix inversion algorithms are $O(N^2)$, where $N$ is the number of local subsystems of the interconnected system. Here we have taken into account that the sparsity of the fundamental matrices can be exploited to decrease the computational complexity of matrix inversion algorithms from $O(N^3)$ to $O(N^2)$. Even if it would be possible to explicitly compute inverses of fundamental matrices, the real-time implementation of the estimators and controllers might be infeasible. Namely, for real-time implementation it is necessary to have a powerful control hardware that can perform $O(N^2)$ complexity calculations (vector-matrix multiplications) during a short time interval limited by a sampling period of a control system.  
\par
On the other hand, control and estimation algorithms, such as the ILC or MHE algorithms, can also be implemented without explicitly computing the inverses of the fundamental matrices. For example, a control action or a state estimate can be efficiently computed using iterative methods for solving large-scale systems of linear equations, such as the Conjugate Gradient (CG) method \cite{saad2003iterative}. However, in practice, the convergence time of any iterative algorithm is limited by the sampling period of a control system. This implies that in practice estimates and control actions are computed using a finite number of iterations. If the number of iterators is small, then it is very-well known that the performance of estimators and controllers can be significantly degraded \cite{khan2008distributing}. On the other hand, computation of estimates (control actions) using iterative methods might not be feasible in large-scale estimation (control) problems that require high sampling frequencies. Another option for the efficient implementation of control algorithms is to (approximately) solve large-scale systems of linear equations using the (incomplete) $LU$ factorization or the Cholesky factorization. 
In spite of the fact that for sparse linear systems of equations it is possible to find sparse $L$ and $U$ factors \cite{saad2003iterative}, the local estimate computed using the $LU$ factorization is a linear combination of all the local data in the network.
\par
The communication and computational problems imposed by the large-scale nature of dynamical networks, motivated the development of distributed/decentralized estimation (control) techniques that can be implemented on a network of local sensors and actuators that communicate locally \cite{massioni2009,pakazad2014distributed,matni2014low,khan2008distributing,motee2008,justin2010distributed,dandrea2003,keviczky2006decentralized,bamieh02distributedcontrol,yo0,Pasqualetti2014,summers2013optimal,dankers2012dynamic}. In recent years, the consensus algorithms \cite{garin2010survey,Olfati2005} have been widely used for the development of distributed estimators and controllers. The consensus algorithms belong to the class of iterative algorithms that in the general case need an infinite number of iterations to converge \cite{khan2008distributing,garin2010survey}. However, like we explained previously, iterative algorithms might not be suitable for large-scale estimation and control problems that require high sampling frequencies.
\par
By analyzing the so-called "lifted state-space representations" of large-scale interconnected systems \cite{verhaegen2007,Haber:13mhe,Haber:14subspace}, one can easily come to the following conclusion. \textit{If the fundamental matrices could be approximated by sparse matrices, then it could be possible to develop novel methods for the identification, estimation and control of large-scale interconnected systems}. For example, in \cite{Haber:13mhe,Haber:14subspace} we have shown that the inverses of sparse, banded Gramians are off-diagonally decaying matrices and that they can be approximated by sparse banded matrices. Using this result we developed novel, distributed/decentralized identification and estimation algorithms for large-scale systems described by sparse banded state-space matrices. These methods identify (estimate) the state-space model of a local subsystem by using only local input-output data. Numerical techniques developed in \cite{Haber:13mhe,Haber:14subspace} can also be used for distributed control of large-scale systems. However, the generalization of the techniques proposed in \cite{Haber:13mhe,Haber:14subspace} for systems described by state-space models whose matrices have arbitrary sparsity patterns (that is, for systems with arbitrary interconnection patterns), is still an open problem. \\
More recently, the problem of designing (sparse) structured feedback gains for control of finite-dimensional interconnected systems has been studied in \cite{linfarjovTAC13admm,linfarjovTAC11al,schuler2011}. Unfortunately, the control design methodologies proposed in these papers are computationally infeasible for large-scale systems, mainly because their computational and memory complexity are at least $O(N^{3})$ and $O(N^2)$, respectively.
\par
In this paper we show that inverses of well-conditioned fundamental matrices can be approximated by sparse matrices.  The approximation methodology established in this paper opens the door to the development of novel methods for distributed estimation, identification and control of large-scale interconnected systems. The novel estimators (controllers) compute local estimates (control actions) simply as linear combinations of inputs and outputs (states) of local subsystems. The size of these local data sets essentially depends on the condition number of the finite-time observability (controllability) Gramian. Furthermore, the developed theory shows that the sparsity patterns of the system matrices of the distributed estimators (controllers) are primarily determined by the sparsity patterns of state-space matrices of large-scale systems. The computational and memory complexity of the approximation algorithms are $O(N)$. Consequently, the proposed approximation methodology is computationally feasible for interconnected systems with an extremely large number of local subsystems.
\par
This paper is organized as follows. In Section \ref{sectionGramians} we explain the importance of finding sparse approximate inverses of fundamental matrices. In Section \ref{ApproximationSection} we present the approximation algorithms. In Section \ref{graphInterpretation} we analyze the sparsity patterns of system matrices of a distributed estimator derived using the approximation algorithms. In Section \ref{numericalSumulations} we present numerical simulations. In Section \ref{conclusionSection} we present conclusions and we discuss future work.
\section{Problem formulation}
In this section we define the class of interconnected systems described by sparse state-space matrices and explain the importance of finding sparse approximate inverses of the fundamental matrices.
\label{sectionGramians}
\subsection{Notation and preliminaries}
The notation $X=[x_{i,j}]$ denotes a matrix whose $(i,j)$ entry is $x_{i,j}$, whereas  $X=[X_{i,j}]$ denotes a block matrix whose $(i,j)$ entry is the matrix $X_{i,j}$. 
The notation $\mathbf{z}=\text{col}\left(\mathbf{z}_{1},\mathbf{z}_{2},\ldots,\mathbf{z}_{M} \right)$ denotes a column vector:  
$\mathbf{z}=[\mathbf{z}_{1}^{T} \;\mathbf{z}_{2}^{T}\; \ldots \; \mathbf{z}_{M}^{T} ]^{T}$. Next, $X=\text{diag}\left(X_{1},X_{2},\ldots,X_{M}\right)$ denotes a block diagonal matrix, with the matrices $X_{1},X_{2},\ldots,X_{M}$ on the main diagonal and $\underline{0}$ denotes a zero matrix whose dimensions will be clear from the context.
\\
We consider large-scale systems consisting of the interconnection of local dynamical subsystems. The interconnection pattern of local subsystems is defined by a \textit{directed graph} $\mathbb{G}=(V,E)$, where $V=\{1,\ldots,N\}$ is a set of \textit{vertices} and $E\subseteq V\times V$ is a set of \textit{edges}. The set $V$ will be also referred to as the \textit{spatial domain}. We assume that $N$ is a large number. For example, $N$ can be in the order of $10^6$ or even larger. Each vertex represents a \textit{local subsystem} $\mathcal{S}_{i}$. The \textit{local state} of $\mathcal{S}_{i}$ is denoted by $\mathbf{x}_{i}(k)\in \mathbb{R}^{n}$. With each edge $(i,j)\in E$ we associate a non-zero matrix $A_{i,j} \in \mathbb{R}^{n\times n}$. The state-space model of $\mathcal{S}_{i}$ has the following form:  
\begin{align}
\mathcal{S}_{i} \left\{ \begin{array} {rll}
\mathbf{x}_{i}(k+1)&=A_{i,i}\mathbf{x}_{i}(k)+\sum_{j \in \mathbb{M}_{i}}^{\;}A_{i,j}\mathbf{x}_{j}(k)+B_{i}\mathbf{u}_{i}(k) \\ 
\mathbf{y}_{i}(k)&=C_{i}\mathbf{x}_{i}(k)+D_{i}\mathbf{u}_{i}(k)+ \mathbf{n}_{i}(k)   \end{array} \right.
\label{localSubSys}
\end{align}
where $\mathbf{u}_{i}(k)\in \mathbb{R}^{m}$ is the \textit{local input}, $\mathbf{y}_{i}(k)\in \mathbb{R}^{r}$ is the \textit{local output}, $\mathbf{n}_{i}(k)\in \mathbb{R}^{r}$ is \textit{the local measurement noise} and $\mathbb{M}_{i}$ is a set of indices defined by 
$\mathbb{M}_{i}=\{j\;| (i,j)\in E,\; j\ne i   \}$. Each index $j\in \mathbb{M}_{i}$ corresponds to a local subsystem $\mathcal{S}_{j}$. That is, the set $\mathbb{M}_{i}$ corresponds to all subsystems that are interconnected with $\mathcal{S}_{i}$. The state-space model of the \textit{global system} (\textit{dynamical network}) is:
\begin{align}
\mathcal{S} \left\{ \begin{array} {rl}
\underline{\mathbf{x}}(k+1)&=\underline{A}\underline{\mathbf{x}}(k)+\underline{B}\underline{\mathbf{u}}(k) \\
\underline{\mathbf{y}}(k)&=\underline{C}\underline{\mathbf{x}}(k)+\underline{D}\underline{\mathbf{u}}(k)+\underline{\mathbf{n}}(k)\end{array} \right. \label{globalSys}
\end{align}
\begin{small}
\begin{align}
&\underline{B}= \text{diag} \left(B_{1}, B_{2},\ldots, B_{N} \right), \underline{C}= \text{diag}\left( C_{1},C_{2},\ldots, C_{N}\right) \notag \\  &\underline{D}= \text{diag}\left( D_{1},D_{2},\ldots, D_{N}\right) \notag \\
& \underline{\mathbf{y}}(k)=\text{col}\left( \mathbf{y}_{1}(k),\ldots,\mathbf{y}_{N}(k) \right), \underline{\mathbf{x}}(k)=\text{col}\left(\mathbf{x}_{1}(k), \ldots, \mathbf{x}_{N}(k)\right),\notag \\
 &\underline{\mathbf{u}}(k)=\text{col}\left(\mathbf{u}_{1}(k), \ldots , \mathbf{u}_{N}(k) \right), \underline{\mathbf{n}}(k)=\text{col}\left(\mathbf{n}_{1}(k),  \ldots,\mathbf{n}_{N}(k)\right)
\notag
\end{align}
\end{small}
and $\underline{A}$ is a matrix whose $(i,j)$ block is $A_{i,j}$. The matrices $\underline{A}\in \mathbb{R}^{Nn\times Nn}$, $\underline{B}\in \mathbb{R}^{Nn\times Nm}$, $\underline{C}\in \mathbb{R}^{Nr\times Nn}$ and $\underline{D}\in \mathbb{R}^{Nr\times Nm}$ are called the \textit{global system matrices}. The vectors $\underline{\mathbf{x}}(k)\in \mathbb{R}^{Nn}$ and $\underline{\mathbf{u}}(k)\in \mathbb{R}^{Nm}$ are called the \textit{global state} and \textit{global input}, respectively. Similarly, the vectors $\underline{\mathbf{y}}(k)\in \mathbb{R}^{Nr}$ and $\underline{\mathbf{n}}(k)\in \mathbb{R}^{Nr}$ are called the \textit{global output} and \textit{global measurement noise}. We assume that $\underline{A}$ is a sparse matrix. That is, we assume that the set $\mathbb{M}_{i}$ has a small number of elements.
\subsection{Lifted system representation}
\label{liftedRepresentationSection}
In a large variety of estimation and control problems, such as the MHE problems \cite{alessandri2003,Haber:13mhe} or ILC problems \cite{Bristow:2006}, the global state-space model \eqref{globalSys} is lifted over the time domain. More precisely, starting from the time instant $k-p$ and by lifting \eqref{globalSys} $p$ time steps, we obtain:
\begin{align}
\underline{\mathbf{x}}(k)&=\underline{A}^{p}\underline{\mathbf{x}}(k-p)+R_{p}\mathbf{U}_{k-p}^{k} \label{liftedClassicalState} \\
\mathbf{Y}_{k-p}^{k}&=O_{p}\underline{\mathbf{x}}(k-p)+\Gamma_{p} \mathbf{U}_{k-p}^{k}+\mathbf{N}_{k-p}^{k} \label{liftedClassicalOutput}
\end{align}
\vspace{-4mm}
\begin{small}
\begin{align}
& \mathbf{U}_{k-p}^{k}=\text{col}\left(\underline{\mathbf{u}}(k-p),..., \underline{\mathbf{u}}(k) \right),  \mathbf{Y}_{k-p}^{k}=\text{col}\left(\underline{\mathbf{y}}(k-p),..., \underline{\mathbf{y}}(k) \right), \notag \\
& \mathbf{N}_{k-p}^{k}=\text{col}\left(\underline{\mathbf{n}}(k-p),..., \underline{\mathbf{n}}(k) \right) \notag 
\end{align}
\end{small}
where $p\ll N$ is the \textit{lifting window}. The matrix $O_{p}\in\mathbb{R}^{N(p+1)r\times Nn}$ is the \textit{p-steps observability matrix}. The matrices $\Gamma_{p}\in\mathbb{R}^{Nn\times N(p+1)m }$ and $R_{p}\in\mathbb{R}^{N(p+1)r \times N(p+1)m}$ are the \textit{p-steps impulse response matrix} and \textit{p-steps controllability matrix}, respectively. The lifted equation \eqref{liftedClassicalOutput} has an elegant graph theoretic interpretation, that will be given in Section \ref{graphInterpretation} (see Remark \ref{remarkGraphTheory}). \textit{Because $\underline{A}$ is sparse and because $p\ll N$, all the matrices in \eqref{liftedClassicalState} and \eqref{liftedClassicalOutput} are sparse.}
\begin{defn} \cite{luenberger1971,Haber:14subspace} The observability index, of the observable global system \eqref{globalSys}, is the smallest integer $\nu$, such that the $\nu$-steps observability matrix $O_{\nu}$ has full column rank, that is $\text{rank}\left(O_{\nu} \right)=nN$ $\hfill \square$
\label{obsIndexDefn}
\end{defn}
Similarly, we can define \textit{the controllability index} as the smallest integer $\theta$, such that the $\theta$-steps controllability matrix $R_{\theta}$ has full rank. We assume that $\nu\ll N$ and $\theta\ll N$. 
\par
The lifted output equation \eqref{liftedClassicalOutput} is formed by lifting the global output and input vectors over the time domain. \textit{However, as it has been shown in \cite{Haber:13mhe,Haber:14subspace}, a more suitable lifting approach for large-scale interconnected systems is to first lift the local outputs and inputs over the time domain and then to lift these lifted vectors over the spatial domain}. To formulate this \textit{structure preserving lifting technique}, we introduce the following notation \cite{Haber:13mhe,Haber:14subspace}. The column vector $\mathcal{Y}_{i,k-p}^{k}\in \mathbb{R}^{(p+1)r}$ is defined by lifting the local output of $\mathcal{S}_{i}$ over the interval $[k-p,k]$:
\begin{align}
 \mathcal{Y}_{i,k-p}^{k}=\text{col}(\mathbf{y}_{i}(k-p),\mathbf{y}_{i}(k-p+1),\ldots,\mathbf{y}_{i}(k))
\label{liftedNotationExplained1}
\end{align}
In the same manner we define the lifted input vector $\mathcal{U}_{i,k-p}^{k}\in \mathbb{R}^{(p+1)m}$ and the lifted measurement noise vector $\mathcal{N}_{i,k-p}^{k}\in \mathbb{R}^{(p+1)r}$. A column vector $\mathcal{Y}_{k-p}^{k}\in \mathbb{R}^{N(p+1)r}$ is defined by lifting lifted local outputs over the spatial domain: $\mathcal{Y}_{k-p}^{k}=\text{col}(\mathcal{Y}_{1,k-p}^{k},\ldots,\mathcal{Y}_{N,k-p}^{k})$.
In the same manner we define the vectors $\mathcal{U}_{k-p}^{k}\in \mathbb{R}^{N(p+1)m}$ and $\mathcal{N}_{k-p}^{k}\in \mathbb{R}^{N(p+1)r}$. It is easy to prove that:
\begin{align}
\mathcal{Y}_{k-p}^{k}=P_{Y}\mathbf{Y}_{k-p}^{k},\;\;\;\mathcal{N}_{k-p}^{k}=P_{Y}\mathbf{N}_{k-p}^{k},\;\;\; \mathcal{U}_{k-p}^{k}=P_{U}\mathbf{U}_{k-p}^{k}
\label{permutationMatrices}
\end{align}
where $P_{Y}$ and $P_{U}$ are permutation matrices. By multiplying the lifted equation \eqref{liftedClassicalOutput} from left with $P_{Y}$ and keeping in mind that permutation matrices are orthogonal, we obtain:
\begin{align} 
\mathcal{Y}_{k-p}^{k}=\mathcal{O}_{p}\underline{\mathbf{x}}(k-p)+\mathcal{G}_{p}\mathcal{U}_{k-p}^{k}+\mathcal{N}_{k-p}^{k}
\label{liftedDataEq}
\end{align}
where the matrices $\mathcal{O}_{p} \in \mathbb{R}^{N(p+1)r\times Nn}$ and $\mathcal{G}_{p}\in \mathbb{R}^{N(p+1)r\times Npm}$ are defined by:
\begin{align}
\mathcal{O}_{p}=P_{Y}O_{p},\;\; \mathcal{G}_{p}=P_{Y}\Gamma_{p}P_{U}^{T}
\label{explanationLiftedDataEq}
\end{align}
On the other hand, using the orthogonality of the permutation matrix $P_{U}$, from \eqref{liftedClassicalState} we obtain:
\begin{align}
\underline{\mathbf{x}}(k)=\underline{A}^{p}\underline{\mathbf{x}}(k-p)+\mathcal{R}_{p}\mathcal{U}_{k-p}^{k} \label{liftedState}
\end{align}
where the matrix $\mathcal{R}_{p} \in \mathbb{R}^{Nn\times N(p+1)m}$ is defined by:
\begin{align}
\mathcal{R}_{p}=R_{p}P_{U}^{T}
 \label{explanationLiftedState}
\end{align}
\textit{Because the matrices $O_{p}$, $R_{p}$ and $\Gamma_{p}$ are sparse, the matrices  $\mathcal{O}_{p}$, $\mathcal{R}_{p}$ and $\mathcal{G}_{p}$ are also sparse.} 
The \textit{finite-time observability Gramian} is given by \cite{lim1996hankel,antoulas2005approximation}:
\begin{align}
\mathcal{W}=\sum_{i=0}^{p}\left(\underline{A}^{T} \right)^{i}\underline{C}^{T}\underline{C} \underline{A}^{i}=O_{p}^{T} O_{p}
\label{obsGrammian}
\end{align}
On the other hand, the \textit{finite-time controllability Gramian} is defined as follows \cite{lim1996hankel,antoulas2005approximation}:
\begin{align}
\mathcal{Q}=\sum_{i=0}^{p}\underline{A}^{i}\underline{B}\underline{B}^{T} \left(\underline{A}^{T} \right)^{i}=R_{p} R_{p}^{T}
\label{conGrammian}
\end{align}
Because permutation matrices are orthogonal, from \eqref{explanationLiftedDataEq} and \eqref{obsGrammian} we have:
\begin{align}
\mathcal{W}=O_{p}^{T} O_{p}=\mathcal{O}_{p}^{T}P_{Y}P_{Y}^{T}\mathcal{O}_{p}=\mathcal{O}_{p}^{T}\mathcal{O}_{p}
\label{ExpressionObservabilityGramian}
\end{align}
Similarly, from \eqref{explanationLiftedState} and \eqref{conGrammian} we have:
 \begin{align}
\mathcal{Q}=\mathcal{R}_{p}\mathcal{R}_{p}^{T}
\label{ExpressionObservabilityGramian2}
\end{align}
Now that we have introduced the lifted system representation and defined finite-time Gramians, we can explain the importance of finding sparse approximate inverses of the fundamental matrices. 
\subsection{Motivation for finding sparse approximate inverses of fundamental matrices}
\label{motivation}
For presentation clarity, the importance of finding sparse approximate inverses of fundamental matrices will be explained on simplified estimation, identification and control problems. Similar conclusions about the importance of finding sparse approximate inverses of fundamental matrices can be drawn from more complex estimation and control problems. Let $\mathcal{O}_{p}=[O_{i,j}]$ and $\mathcal{G}_{p}=[G_{i,j}]$, where $O_{i,j}\in \mathbb{R}^{(p+1)r\times n}$ and $G_{i,j} \in \mathbb{R}^{n \times (p+1)m}$. Then, the $i$th block equation of \eqref{liftedDataEq} can be written as follows:
\begin{align}
\mathcal{Y}_{i,k-p}^{k}=\sum_{j\in \mathbb{M}_{O,i}}O_{i,j} \mathbf{x}_{j}(k-p)+\sum_{j\in \mathbb{M}_{G,i}}G_{i,j}\mathcal{U}_{j,k-p}^{k}+\mathcal{N}_{i,k-p}^{k}
\label{outputLocalSum}
\end{align}
where $\mathbb{M}_{O,i}=\{j \;|\; O_{i,j }\neq \underline{0} \}$ and $\mathbb{M}_{G,i}=\{j \; |\; G_{i,j}\neq \underline{0} \}$ are sets of indices.  
\subsubsection{Estimation and identification problems}
Because the matrices $\mathcal{O}_{p}$ and $\mathcal{G}_{p}$ are sparse, from \eqref{outputLocalSum} we have that the lifted local output $\mathcal{Y}_{i,k-p}^{k}$ depends on a relatively few local states and inputs. \textit{However, to estimate $\mathbf{x}_{i}(k-p)$ in the least-squares sense, the local subsystem $\mathcal{S}_{i}$ needs to take into account \textbf{all} local inputs and outputs in the network.} To show this, let's take a look at the lifted data equation \eqref{liftedDataEq}. The global state vector can be estimated by solving the following least-squares problem:
\begin{align}
\min_{\underline{\mathbf{x}}(k-p)} \left\| \mathcal{Y}_{k-p}^{k}-\mathcal{G}_{p}\mathcal{U}_{k-p}^{k}- \mathcal{O}_{p}\underline{\mathbf{x}}(k-p) \right\|_{2}^{2}
\label{leastSquares12}
\end{align}
Let us assume that $p\ge \nu$ ($\nu$ is the observability index, see Definition \ref{obsIndexDefn}).
Because $p\ge \nu$, the matrix $O_{p}$ has full column rank. Furthermore, because $\mathcal{O}_{p}=P_{Y} O_{p}$ and because permutation matrices do not change matrix rank, we conclude that $\mathcal{O}_{p}$ has full column rank. This implies that the solution of \eqref{leastSquares12} is given by:
\begin{align}
\hat{\underline{\mathbf{x}}}(k-p)=\mathcal{W}^{-1}\mathcal{O}_{p}^{T}\left(\mathcal{Y}_{k-p}^{k}-\mathcal{G}_{p}\mathcal{U}_{k-p}^{k}\right)
\label{leastSquares1}
\end{align} 
where $\hat{\underline{\mathbf{x}}}(k-p)=\text{col}\left(\hat{\mathbf{x}}_{1}(k-p),...,\hat{\mathbf{x}}_{N}(k-p) \right)$. Although $\mathcal{O}_{p}$ is sparse, the matrix $\mathcal{W}^{-1}$ is fully populated (inversion "destroys" the matrix structure). This means that the local state estimate $\hat{\mathbf{x}}_{i}(k-p)$ is a linear combination of all local (lifted) inputs and outputs in the network. \textit{That is, to calculate $\hat{\mathbf{x}}_{i}(k-p)$, the local subsystem $\mathcal{S}_{i}$ needs to know the input-output data of \textbf{all} local subsystems in the network.} Because of the various communication constraints that are present in practice, $\mathcal{S}_{i}$ cannot obtain the input-output data of all local subsystems in the network. Consequently, it might not be possible to compute $\hat{\mathbf{x}}_{i}(k-p)$ as a linear combination of all local data in the network. Furthermore, because of the large-scale nature of the network, it might not be possible to explicitly compute and store $\mathcal{W}^{-1}$.
\par
Using the results of \cite{Haber:14subspace}, from \eqref{liftedDataEq} and \eqref{liftedState} is easy to derive the following AutoRegressive eXogenous (ARX) \cite{verhaegen2007} model:
\begin{align}
  \underline{\mathbf{y}}(k)=&\underline{C}\underline{A}^{p}\mathcal{W}^{-1}\left(\mathcal{O}_{p}^{T}\mathcal{Y}_{k-p}^{k}-\mathcal{O}_{p}^{T}\mathcal{G}_{p}\mathcal{U}_{k-p}^{k}\right)+\underline{C}\mathcal{R}_{p}\mathcal{U}_{k-p}^{k}
\label{ARXmodel}
\end{align}
where we neglect the measurement noise vector. From the $i$th block equation of the ARX model \eqref{ARXmodel}, we can derive a local ARX model for $\mathbf{y}_{i}(k)$. The identification problem consists of estimating the parameters of the local ARX model using the sequence of the local input-output data. Because $\mathcal{W}^{-1}$ is dense, we see that to identify the parameters of the local ARX model we need to take inputs and outputs of all local subsystems in the network. \subsubsection{Control problems}
Similar difficulties appear in control problems for large-scale systems. To illustrate this, let's look at the lifted state equation \eqref{liftedState}. Because $\underline{A}$ is sparse and $p\ll N$, the matrices $\underline{A}^{p}$ and $\mathcal{R}_{p}$ are also sparse. Because of this, from \eqref{liftedState} we conclude that $\mathbf{x}_{i}(k)$ is a linear combination of past local states and 
past local inputs of systems that are in some neighborhood of $\mathcal{S}_{i}$ (this neighborhood is determined by the sparsity patterns of $\underline{A}^{p}$ and $\mathcal{R}_{p}$, for more details see Section \ref{graphInterpretation}). \textit{However the input sequence $\mathcal{U}_{i,k-p}^{k}$ that takes $\mathcal{S}_{i}$ from $\mathbf{x}_{i}(k-p)$ to $\mathbf{x}_{i}(k)$ is a linear combination of the states of all the local subsystems in the network.} To show this, consider the system of equations \eqref{liftedState} that we want to solve for $\mathcal{U}_{k-p}^{k}$. Because the system of equations \eqref{liftedState} is under-determined, we are searching for the \textit{least-norm solution}\footnote{In fact, the solution of \eqref{leastNorm} minimizes the input energy.}:
\begin{align}
&\text{minimize} \;\;\; \left\|\mathcal{U}_{k-p}^{k}  \right\|_{2}^{2} \label{leastNorm} \\
&\text{subject to \;\;\; \eqref{liftedState}} \notag
\end{align}
Assume that $p\ge \theta$, where $\theta$ is the controllability index. Under this condition the matrix $\mathcal{R}_{p}$ has full rank, and the solution of \eqref{leastNorm} is given by:
\begin{align}
\mathcal{U}_{k-p}^{k}= \mathcal{R}_{p}^{T}\mathcal{Q}^{-1} \left( \underline{\mathbf{x}}(k)-\underline{A}^{p}\underline{\mathbf{x}}(k-p)\right) \label{liftedStateSolved}
\end{align}
Because the matrix $\left(\mathcal{R}_{p}\mathcal{R}_{p}^{T}\right)^{-1}$ is fully populated, from the $i$th block equation of \eqref{liftedStateSolved} we see that the local input sequence $\mathcal{U}_{i,k-p}^{k}$ is a linear combination of all local states in the network. Like it is explained previously, due to various computation and communication constraints that are present in the practice, the local control action cannot be computed as a linear combination of all the data in the network. 
\par
Similar difficulties appear in predictive control problems or in the ILC problems in which the impulse response matrix needs to be inverted. For simplicity, let assume that the initial state $\underline{\mathbf{x}}(k-p)$ and the measurement noise in \eqref{liftedDataEq} are zero. A simplified predictive control problem at the time instant $k-p$, consists of finding the control sequence $\mathcal{U}_{k-p}^{k}$ that will produce the desired output $\mathcal{Y}_{k-p}^{k}$. This can be achieved by solving the following system of equations:
\begin{align}
\mathcal{Y}_{k-p}^{k}=\mathcal{G}_{p}\mathcal{U}_{k-p}^{k}
\label{leastSquares12Input}
\end{align}
Assuming that $\mathcal{G}_{p}$ has full column rank, the solution of \eqref{leastSquares12Input} is:
\begin{align}
\hat{\mathcal{U}}_{k-p}^{k}=\left(\mathcal{G}_{p}^{T}\mathcal{G}_{p} \right)^{-1}\mathcal{G}_{p}^{T}\mathcal{Y}_{k-p}^{k}
\label{leastSquares12InputSolution}
\end{align}
Because in the general case $\left(\mathcal{G}_{p}^{T}\mathcal{G}_{p} \right)^{-1}$ is fully populated, from \eqref{leastSquares12InputSolution} we have that the local input sequence $\hat{\mathcal{U}}_{i,k-p}^{k}$ is a linear combination of all the local lifted outputs in the network.
\par
From \eqref{leastSquares1} we see that if the inverse of the observability Gramian could be approximated by a sparse matrix, then $\hat{\mathbf{x}}_{i}(k-p)$ could be estimated simply as a linear combination of the input-output data of the local subsystems that are in some neighborhood of $\mathcal{S}_{i}$ (this neighborhood would be determined by the sparsity pattern of an approximate, sparse inverse of $\mathcal{W}$ and sparsity patterns of  $\mathcal{O}_{p}^{T}$ and $\mathcal{G}_{p}$, see Section \ref{graphInterpretation} for more details). Similarly to the estimation problem, from \eqref{ARXmodel} we see that if there would exist a sparse approximate inverse of  $\mathcal{W}$, then the local ARX model of $\mathcal{S}_{i}$ would depend on the inputs and outputs of local subsystems that are in some neighborhood of $\mathcal{S}_{i}$. Consequently, the parameters of the local ARX model could be identified in the decentralized manner \cite{Haber:14subspace}.
\par
 On the other hand, from \eqref{liftedStateSolved} we see that if the inverse of the finite-time controllability Gramian could be approximated by a sparse matrix, then the local input sequence $\mathcal{U}_{i,k-p}^{k}$ could be determined simply as a linear combination of states of the local subsystems that are in some neighborhood of $\mathcal{S}_{i}$.  Finally, from \eqref{leastSquares12InputSolution} we see that if the pseudo-inverse of the impulse response matrix could be approximated by a sparse matrix, then the control action $\mathcal{U}_{i,k-p}^{k}$ could be computed simply as a linear combination of the outputs of the local subsystems that are neighbors to $\mathcal{S}_{i}$.
\\ 
\textit{Motivated by these observations, in the sequel we develop a framework for approximating the inverses of the fundamental matrices by sparse matrices.} For brevity and presentation clarity, in this paper we only consider approximation of the finite-time observability Gramian. 
\section{Approximate inverses of finite-time Gramians}  
\label{ApproximationSection}
A "naive" approach for computing the sparse approximate inverses of the finite-time Gramians, consists of first computing the fully populated inverses and then setting to zero their small entries. Because of the $O(N^2)$ computational and memory complexity of the (sparse) matrix inversion algorithms, the naive approach cannot be used in practice. 
In \cite{Haber:13mhe,Haber:14subspace} we have shown that in the case of large-scale systems with banded state-space matrices, the inverses of the finite-time Gramians are off-diagonally decaying matrices\cite{demko1984}. This result is important from computational point of view because in \cite{benzi2007} it has been shown that off-diagonally decaying matrices can be efficiently approximated by sparse matrices. In \cite{Haber:13mhe} we have used the Chebyshev matrix polynomials \cite{benzi2007,Mathar2006} to compute sparse approximate inverses of the finite-time observability Gramians. Another option for computing sparse approximate inverses of sparse matrices is to use the Newton-Schultz iteration \cite{haberThesis} or to use the methods proposed in \cite{chow1998approximate,grote1997}. Due to its simplicity and fast convergence rate, in this paper we use the Newton-Schultz iteration. Because in some cases the Newton-Schultz iteration fails to converge, in Section \ref{other} we briefly explain the main ideas of the approximation methods proposed in \cite{chow1998approximate,grote1997}.
\par
The Newton-Schultz iteration for approximating the inverse of $\mathcal{W}$ is defined by \cite{hackbusch2008approximate,pan1991improved}:
\begin{align}
X_{k+1}=X_{k}\left(2I-\mathcal{W}X_{k}\right), \; k=0,1,2 ,\ldots
\label{NewtonSchultz}
\end{align}
where $X_{k}$ is an approximate inverse at the $k$th iteration. Using the results of \cite{pan1991improved} it is easy to prove the following theorem.
\begin{thm}
Let the initial value for the Newton-Schultz iteration be chosen as follows:
\begin{align}
X_{0}=\frac{2}{a^2+b^2}\mathcal{W}
\label{initialNewton}
\end{align}
where $a$ and $b$ are minimal and maximal singular values of $\mathcal{W}$, respectively. Furthermore, let the approximation accuracy of the Newton-Schultz iteration be quantified by:
\begin{align}
\mathcal{E}_{k}=I-\mathcal{W}X_{k}
\label{approximationError1}
\end{align}
then
\begin{align}
\left\|\mathcal{E}_{k}\right\|_{2}\le \left(\frac{\kappa^2-1}{\kappa^2+1}\right)^{2^{k}}
\label{errorFinalBound}
\end{align}
where $\kappa=b/a$ is the condition number of $\mathcal{W}$.
\label{convergenceNewtonSchultz}
\end{thm}
\textit{Proof} Because $\mathcal{W}$ is a symmetric matrix, its Singular Value Decomposition (SVD) is:
\begin{align}
\mathcal{W}=U\Sigma U^{T}
\label{svd1}
\end{align}
where $U$ is the unitary matrix and $\Sigma$ is a diagonal matrix of singular values. From \eqref{initialNewton}, \eqref{approximationError1} and \eqref{svd1} we have:
\begin{align}
\mathcal{E}_{0}=U\left(I-\frac{2}{a^2+b^2}\Sigma^2 \right)U^T
\label{proofThm1}
\end{align}
It is easy to prove that:
\begin{align}
\left\|\mathcal{E}_{0}\right\|_{2}=\frac{b^2-a^2}{b^2+a^2}=\frac{\kappa^2-1}{\kappa^2+1}
\label{proofThm2}
\end{align}
Next, from \eqref{NewtonSchultz} and \eqref{approximationError1} we have:
\begin{align}
\mathcal{E}_{k+1}=\left(I-\mathcal{W}X_{k}\right)\left(I-\mathcal{W}X_{k}\right)
\label{proofThm3}
\end{align}
From \eqref{proofThm3} we have:
\begin{align}
\left\|\mathcal{E}_{k+1}\right\|_{2}\le \left\| \mathcal{E}_{k}\right\|_{2}^{2}
\label{proofThm4}
\end{align}
\textit{which proves that the Newton-Schultz iteration has a quadratic convergence rate.} From \eqref{proofThm4} we have:
\begin{align}
\left\|\mathcal{E}_{k}\right\|_{2}\le \left\|\mathcal{E}_{0} \right\|_{2}^{2^{k}}
\label{proofThm5}
\end{align}
Substituting \eqref{proofThm2} in \eqref{proofThm5} we obtain \eqref{errorFinalBound}. $\hfill \square$
\par
Because $\underline{A}$ is sparse and because $p\ll N$, the finite-time observability Gramian $\mathcal{W}$ is sparse. Due to this, the constants $a$ and $b$ can be computed with $O(Nn)$ complexity \cite{Haber:13mhe}.
\par
From \eqref{initialNewton} we have that the initial guess for the Newton-Schultz iteration $X_{0}$, has the same sparsity pattern as $\mathcal{W}$. However, from \eqref{NewtonSchultz} we see that $X_{1}$ has a larger number of non-zero elements than $X_{0}$. As the number of iterations $k$ increase, the matrix $X_{k}$ becomes denser and denser. Direct consequence of this is that the computational and memory complexity of the Newton-Schultz iteration increase with $k$. Furthermore, for relatively large $k$, the matrix $X_{k}$ becomes fully populated. 
\\
However, Theorem \ref{convergenceNewtonSchultz} shows that if $\mathcal{W}$ is well-conditioned (that is, $\kappa$ is close to $1$), then we need a relatively small number of iterations $k$ to accurately approximate $\mathcal{W}^{-1}$. Namely, if $\kappa$ is close to 1, then there exists small $k$ for which the upper bound on the right-hand side of \eqref{errorFinalBound} is close to zero. \textit{This means that if the matrix $\mathcal{W}$ is well conditioned, then its inverse can be accurately approximated by a sparse matrix}. We also conclude that for a fixed approximation accuracy, the matrix $\mathcal{W}$ with a larger condition number has the approximate inverse with a larger number of non-zero elements (because for a matrix with a larger condition number, we  need a larger number of iterations to achieve the same accuracy). In Section \ref{graphInterpretation} we give a detailed analysis of the sparsity structure of $\mathcal{W}$.
\par
In many estimation and control problems, regularization parameters are added to finite-time Gramians. By increasing the value of the regularization parameter, the condition number of the regularized Gramian can be decreased. \textit{This implies that regularization parameters can be used to reduce the number of non-zero elements of the approximate inverses of the fundamental matrices.} For example, the least squares problem \eqref{leastSquares12} can be modified by penalizing the global state: 
\begin{align}
\min_{\underline{\mathbf{x}}(k-p)} \left\{\left\| \mathcal{Y}_{k-p}^{k}-\mathcal{G}_{p}\mathcal{U}_{k-p}^{k}- \mathcal{O}_{p}\underline{\mathbf{x}}(k-p) \right\|_{2}^{2}+\mu \left\|\underline{\mathbf{x}}(k-p)\right\|_{2}^{2} \right\}
\label{leastSquares14}
\end{align}
where $\mu$ is the regularization parameter. Let $\mathcal{W}_{r}=\mathcal{W}+\mu I $. It can be easily shown that the solution of \eqref{leastSquares14} is defined by replacing $\mathcal{W}^{-1}$ with $\mathcal{W}_{r}^{-1}$ in \eqref{leastSquares1}. The condition number of $\mathcal{W}_{r}$ is $\left(b+\mu \right) /\left(a+\mu \right) $. By increasing $\mu$ the condition number of $\mathcal{W}_{r}$ can be decreased. This way, we can ensure that even if $\mathcal{W}$ is badly conditioned, its regularized inverse $\mathcal{W}_{r}^{-1}$ can still be accurately approximated by a sparse matrix using only a few Newton-Schultz iterations. 
\subsection{Sparsification methods}
 To ensure that each Newton-Schultz iteration can be computed with linear computational and memory complexity, and to guarantee that the approximate inverses are sparse, we need to sparsify $X_{k}$ after each iteration. This can be achieved by modifying the Newton-Schultz iteration as follows:
\begin{align}
Z_{k}=X_{k}\left(2I-\mathcal{W}X_{k}\right), \;\;
X_{k+1}=H\left(Z_{k}\right), k=0,1,2,\ldots
\label{NewtonSchultzDropping}
\end{align}
where $H\left( Z_{k} \right)$ is a sparsification operator. There are two types of sparsification operators. The first operator sets to zero all elements of $Z_{k}$ that are outside some prescribed sparsity pattern. To be more precise, let $P\in \mathbb{R}^{Nn\times Nn}$ be a binary matrix describing the \textit{desired sparsity pattern} of the approximate inverse. Furthermore, let $P=[p_{i,j}]$ and $Z_{k}=[z_{i,j}]$. The first sparsification operator $H_{1}\left(Z_{k}\right)$ is defined element-wise by:
\begin{align}
\left(H_{1}\right)_{i,j}=\left\{ \begin{array} {rl}
z_{i,j}, &   \;\;  p_{i,j}= 1  \\
0, &  \;\; p_{i,j}=0 \end{array} \right. 
\label{eq:operator1}
\end{align}
where $\left(H_{1}\right)_{i,j}$ is the $(i,j)$ entry of the sparsification operator  $H_{1}$. The second sparsification operator, denoted by $H_{2}$, neglects "small" elements of $Z_{k}$. It is defined element-wise by:
\begin{align}
\left( H_{2} \right)_{i,j}=\left\{ \begin{array} {rl}
z_{i,j}, &  \;\;  |z_{i,j}|> \phi  \\
0, &   \; |z_{i,j}| \le \phi \end{array} \right. 
\label{eq:operator1}
\end{align}
where $\phi$ is a \textit{dropping parameter} and $| \cdot|$ is the absolute value. Provided that the sparsity pattern $P$ or dropping parameter $\phi$ are properly chosen, in \cite{hackbusch2008approximate} it has been shown that the Newton-Schultz iteration is robust with respect to (small) errors introduced by the sparsification operators. The fastest algorithms are obtained by combining both of the sparsification operators. 

\subsection{How to chose the parameters of the sparsification operators?}
\label{SparsityPatternSelection}
If $\underline{A}$ is sparse and banded, then $\mathcal{W}$ is also sparse and banded. Inverses of banded observability Gramians are off-diagonally decaying matrices \cite{Haber:13mhe,Haber:14subspace,haberThesis}. Moreover, the rate of the off-diagonal decay of $\mathcal{W}^{-1}$ depends on the condition number of $\mathcal{W}$ \cite{demko1984}. This implies that in the case of approximating inverses of banded observability Gramians, the pattern matrix $P$ should be a banded matrix. The bandwidth of $P$ should be selected on the basis of the condition number of $\mathcal{W}$. This bandwidth can be selected using the equation 17. in \cite{Haber:14subspace}.
\par
In the general case, when the matrix $\underline{A}$ is sparse but it does not have a specific structure, the sparsity pattern of $P$ can be formed by summing up the powers of $\mathcal{W}$ \cite{huckle1999apriori}.  Namely, consider the Neumann series:
\begin{align}
(I-Y)^{-1}=\sum_{i=0}^{\infty} Y^{i}
\label{NeumannExpansion}
\end{align}
where $Y$ is a matrix satisfying $\left\| Y \right\|_{2}<1$. Let $Y=I-\epsilon\mathcal{W}$, where $\epsilon$ is a small number. The parameter $\epsilon$ ensures that the singular values of $\mathcal{W}$ are scaled such that $\left\| I-\epsilon \mathcal{W} \right\|_{2}<1$. From \eqref{NeumannExpansion} we have:
\begin{align}
\mathcal{W}^{-1}=\epsilon \sum_{i=0}^{\infty}\left(I-\epsilon \mathcal{W} \right)^{i}
\label{NeumannExpansion2}
\end{align}
Assuming that $\epsilon \left\| \left(I-\epsilon \mathcal{W} \right)^{i}\right\|_{2}$ is small for $i\ll N$, from \eqref{NeumannExpansion2} we have that a relatively good estimate of the sparsity pattern of $\mathcal{W}^{-1}$ is given by the sum of the first $i-1$ terms in \eqref{NeumannExpansion2}. Because the largest sparsity pattern of $\left(I-\epsilon \mathcal{W} \right)^{i}$ is determined by the sparsity pattern of $\mathcal{W}^{i}$, we see that a relatively good estimate of the sparsity pattern of $\mathcal{W}^{-1}$ is determined by a matrix obtained by summing the powers of $\mathcal{W}$. Further details on the choice of the a priori sparsity patterns of approximate inverses can be found in \cite{tang1999toward,huckle1999apriori}.
\par
In some cases the matrix $\mathcal{W}$ can be sparsified and consequently, its approximate inverse can be additionally sparsified. Namely, if the global system is asymptotically stable, then $\underline{A}^{s}\approx \underline{0}, \; \left( \underline{A}^{T}\right)^{s}\approx \underline{0}$, for some sufficiently large $s$. If there exists $s<p $, then the matrix $\mathcal{W}$ can be sparsified by neglecting all the powers of $\underline{A}$ and $\underline{A}^{T}$ that are larger than $s$ (see the definition of the finite-time observability Gramian \eqref{obsGrammian}). Using this idea we can also significantly sparsify the powers of $\mathcal{W}$.
\par
Our experience with computing approximate inverses of ill-conditioned Gramians shows that the Newton-Schultz iteration can diverge. For ill-conditioned Gramians, this occurs if the pattern matrix $P$ is too sparse, or if the truncation parameter is relatively high. One of the ways to stabilize the Newton-Schultz iteration is to increase the number of non-zero elements of $P$ or to decrease $\phi$. The price of this is increased computational complexity. Another way for overcoming this problem is to decrease the condition number of $\mathcal{W}$ by introducing the regularization parameter. However, in some cases there might not be a physical justification for introducing regularization parameters.
\subsection{Other approaches for computing sparse approximate inverses}
\label{other}
Luckily, there are other methods for computing sparse approximate inverses \cite{chow1998approximate,grote1997}. These methods are not based on the Newton-Schultz iteration and they might be more appropriate in the case of ill-conditioned Gramians. On the other hand, the computational complexity of these methods might be higher (except in the case of parallel implementation). The main idea of these methods is to find an approximate, sparse inverse of $\mathcal{W}$ by solving: 
\begin{align}
\min_{X} \left\|I-\mathcal{W}X\right\|_{F}^{2}
\label{sparseApproximateInverseCostFunction}
\end{align}
where $X\in \mathbb{R}^{Nn\times Nn}$ is a sparse approximate inverse of $\mathcal{W}$ that we want to find and $\left\| \cdot \right\|_{F}$ is the Frobenius norm. It can be easily shown that 
\begin{align}
\left\|I-\mathcal{W}X\right\|_{F}^{2}=\sum_{j=1}^{Nn}\left\|\mathbf{t}_{j}-\mathcal{W}\mathbf{v}_{j}  \right\|_{2}^{2}
\label{sparseApproximateInverseCostFunction2}
\end{align}
where $\mathbf{t}_{j}$ and $\mathbf{v}_{j}$ are the $j$th columns of $I$ and $X$, respectively. Because the columns of $X$ are independent, we see that the optimization problem \eqref{sparseApproximateInverseCostFunction} can be solved by solving $Nn$ independent least squares problems:
 \begin{align}
\min_{\mathbf{v}_{j}}\left\|\mathbf{t}_{j}-\mathcal{W}\mathbf{v}_{j}  \right\|_{2}^{2}, \;\; j=1,\ldots,Nn
\label{sparseApproximateInverseCostFunction3}
\end{align}
Because $\mathcal{W}$ and $X$ are sparse, the optimization problems $\eqref{sparseApproximateInverseCostFunction3}$ can be reduced to low-dimensional least-squares problems that can be solved efficiently. Furthermore, the least squares problems \eqref{sparseApproximateInverseCostFunction3} can be solved in parallel. However, the main difficulty is how to chose the sparsity pattern of $X$. The methods proposed in \cite{chow1998approximate,grote1997} chose the sparsity pattern of $X$ iteratively. These methods start with an initial pattern of $X$ and on the basis of a certain criteria they iteratively update the initial sparsity pattern. This process is repeated until a good approximation accuracy has been achieved or a prescribed number of non-zero elements of $X$ has been reached. However, this process can be computationally expensive. An alternative, less computationally demanding approach \cite{huckle1999apriori}, is to a priori chose the sparsity pattern of $X$, and to directly solve \eqref{sparseApproximateInverseCostFunction3}. For example, a priori sparsity pattern of $X$ can be selected using the guidelines given in Section \ref{SparsityPatternSelection}.
\section{The structure of the distributed estimator}
\label{graphInterpretation}
Let the approximate inverse of $\mathcal{W}$, computed using the Newton-Schultz iteration, be denoted by $X$. By substituting $\mathcal{W}^{-1}$ with $X$ in \eqref{leastSquares1}, we obtain:
\begin{align}
\hat{\underline{\mathbf{x}}}(k-p)\approx \underbrace{X\mathcal{O}_{p}^{T}}_{L}\mathcal{Y}_{k-p}^{k}-\underbrace{X\mathcal{O}_{p}^{T}\mathcal{G}_{p}}_{Q}\mathcal{U}_{k-p}^{k}
\label{approximateInverseSolution}
\end{align}
Because $X$, $\mathcal{O}_{p}$ and $\mathcal{G}_{p}$ are sparse, the matrices $L$ and $Q$ are also sparse. Let $L=[L_{i,j}]$ and $Q=[Q_{i,j}]$, where $L_{i,j}\in \mathbb{R}^{n\times (p+1)r}$ and $Q_{i,j} \in \mathbb{R}^{n \times (p+1)m}$. Then from \eqref{approximateInverseSolution} we have:
\begin{align}
\hat{\mathbf{x}}_{i}(k-p)\approx \sum_{j\in \mathbb{M}_{L,i}}L_{i,j} \mathcal{Y}_{j,k-p}^{k}-\sum_{j\in \mathbb{M}_{Q,i}}Q_{i,j}\mathcal{U}_{j,k-p}^{k}
\label{outputLocalSum22}
\end{align}
where $\mathbb{M}_{L,i}=\{j \;|\; L_{i,j }\neq \underline{0} \}$ and $\mathbb{M}_{Q,i}=\{j \; |\; Q_{i,j}\neq \underline{0} \}$ are sets of indices. To compute $\hat{\mathbf{x}}_{i}(k-p)$, the local subsystem $\mathcal{S}_{i}$ needs to receive the outputs of all the local subsystems $\mathcal{S}_{j}$, where $j\in \mathbb{M}_{L,i}$, and the inputs of all the local subsystems $\mathcal{S}_{j}$, where $j\in \mathbb{M}_{Q,i}$. \textit{That is, the equation \eqref{outputLocalSum22} constitutes the distributed state estimator (see Remark \ref{MHEremark}).} The graphs describing the communication links between the local subsystems of the distributed estimator are determined by the sparsity patterns of $L$ and $Q$. In this section we analyze the sparsity patterns of these matrices.
\begin{rem}
For presentation clarity, in this paper we have derived distributed least-squares estimator \eqref{outputLocalSum22}. However, the developed approximation methods can be used to derive a more complex distributed estimators, such as the moving horizon estimators \cite{Haber:13mhe}.  $\hfill \square$
\label{MHEremark}
\end{rem}
\textit{For simplicity and without loss of generality, we assume that all non-zero blocks of $\underline{A}$, $\underline{B}$, $\underline{C}$ and $\underline{D}$ are positive. This way, we ensure that there are no numerical cancellations in matrix expressions involving these matrices. This is a standard assumption in methods that analyze the structure of sparse matrices \cite{davis2006direct}.} 
\par
  Briefly speaking, the sparsity patterns of $L$ and $Q$ are determined by: 1) The sparsity pattern of $\underline{A}$; 2) The lifting window $p$; 3) The number of iterations necessary to compute $X$ and if the sparsified Newton-Schultz iteration is used, by the sparsity pattern of $P$.
\\
 Before we relate the structure of $L$ and $Q$ with the structure of $\underline{A}$, we make the following observation. As it is explained in Section \ref{ApproximationSection}, if $\mathcal{W}$ is well-conditioned, then $X$ can be accurately approximated by a sparse matrix without the need to use the sparsification operators. That is, if $\mathcal{W}$ is well-conditioned then the matrices $L$ and $Q$ are sparse. \textit{This means that if $\mathcal{W}$ is well-conditioned, then we need a small amount of communication between the local subsystems to compute the local state estimate \eqref{outputLocalSum22}}.
\par
 To relate the sparsity pattern of $\underline{A}$ with the patterns of $L$ and $Q$, we need to associate the interconnection matrix (adjacency matrix) with the global system $\mathcal{S}$ \cite{siljak1991}.
\begin{defn} 
The interconnection matrix of the global system $\mathcal{S}$ (the adjacency matrix of the graph $\mathbb{G}$) is a binary $N\times N$ matrix $\overline{A}=[\overline{a}_{i,j}]$ defined element-wise by:
\begin{align}
\overline{a}_{i,j}= \left\{ \begin{array} {rl}
1 &, \;\; (i,j)\in E  \Leftrightarrow 	A_{i,j}\neq \underline{0} \\
0 &, \;\; (i,j)\notin E \Leftrightarrow 	A_{i,j}= \underline{0} \end{array} \right. 
\label{eq:interconnectionMatrix}
\end{align}
 $\hfill \square$
\end{defn}
The matrix $\overline{A}$ describes the block sparsity pattern of $\underline{A}$. Because $\underline{A}$ is a sparse matrix, the matrix $\overline{A}$ is also sparse. The matrix $\underline{C}$ is a block diagonal matrix. Similarly to the definition of the adjacency matrix $\overline{A}$, we introduce an $N\times N$ identity matrix, denoted by $\overline{C}$, that describes the sparsity pattern of $\underline{C}$.
 Next, we introduce a matrix operator that transforms an arbitrary matrix into a binary matrix describing its sparsity pattern.
\begin{defn}
Let $Z=[z_{i,j}]$ be an arbitrary $N\times N$ matrix. An $N\times N$ matrix $T(Z)=[t_{i,j}]$ is defined element-wise by:
\begin{align}
t_{i,j}=\left\{ \begin{array} {rl}
1, &   \;\;  z_{i,j}\neq 0  \\
0, &   \;\; z_{i,j}=0 \end{array} \right. 
\label{eq:interconnectionMatrix}
\end{align}
\label{binaryOperator}
 $\hfill \square$
\end{defn}
The first step in the analysis of the sparsity patterns of $L$ and $Q$ is to analyze the sparsity pattern of $\mathcal{O}_{p}$. The block sparsity pattern of $\mathcal{O}_{p}=[O_{i,j}]$ can be described by an $N\times N$ matrix $\overline{\mathcal{O}}=[o_{i,j}]$, defined element-wise by:
\begin{align}
o_{i,j}= \left\{ \begin{array} {rl}
1, &   \;\;  O_{i,j}\neq \underline{0} \\
0, &   \;\; O_{i,j}=\underline{0} \end{array} \right. 
\label{adjacencyofObs}
\end{align}
This matrix $\mathcal{O}_{p}$ is obtained by permuting the p-steps observability matrix $O_{p}$:
\begin{align}
O_{p}=\begin{bmatrix}\underline{C}^{T} & \left(\underline{C}\underline{A}\right)^{T} & \hdots & \left(\underline{C}\underline{A}^{p}\right)^{T} \end{bmatrix}^{T}
\label{pObservability}
\end{align}
The block sparsity pattern of $\underline{C}\underline{A}^{l}$, where $l=0,1,\ldots, p$, can be described by the sparsity pattern of $\overline{C}\overline{A}^{l}$. Because the matrix $\overline{C}$ is an identity matrix, the sparsity pattern of the $l$th block of \eqref{pObservability} can be described by the sparsity pattern of  $\overline{A}^{l}$. 
From the equation \eqref{liftedClassicalOutput} we have that if the $(i,j)$ element of $\overline{A}^{l}$ is not equal to zero, then the local output $\mathbf{y}_{i}(k-p+l)$ depends on the local state $\mathbf{x}_{j}(k-p)$. This furthermore implies that the lifted output $\mathcal{Y}_{i,k-p}^{k}=\text{col}\left(\mathbf{y}_{i}(k-p),\ldots,\mathbf{y}_{i}(k)\right)$ depends on the local state $\mathbf{x}_{j}(k-p)$, if the $(i,j)$ entry of any of the matrices $I,\overline{A},\overline{A}^2,\ldots, \overline{A}^{p}$ is non-zero. That is, $\mathcal{Y}_{i,k-p}^{k}$ depends on $\mathbf{x}_{j}(k-p)$ if the $(i,j)$ entry of the matrix $I+\overline{A}+\overline{A}^2+\ldots+\overline{A}^{p}$ is non-zero. On the other hand, if $\mathcal{Y}_{i,k-p}^{k}$ is depending on $\mathbf{x}_{j}(k-p)$ then at least one entry of the matrix $O_{i,j}$ in \eqref{outputLocalSum} is non-zero.
This implies that the matrix $O_{i,j}$ has non-zero entries if the $(i,j)$ entry of
$I+\overline{A}+\overline{A}^2+\ldots+\overline{A}^{p}$ is not equal to zero. If the matrix $O_{i,j}$ contains non-zero entries then from \eqref{adjacencyofObs} we have: $o_{i,j}=1$. All this implies that (see Remark \ref{remarkGraphTheory}):
\begin{align}
\overline{\mathcal{O}}=T\left(I+\overline{A}+\overline{A}^2+\ldots+\overline{A}^{p}\right)
\label{matrixPowersSumAdjacencyConclusion}
\end{align}
\begin{rem}
From the graph theory \cite{saad2003iterative} it is very well known that the $(i,j)$ entry of $\overline{A}^{l}$ is the number of paths of length $l$ from the vertex $i$ to the vertex $j$ in $\mathbb{G}$. \textit{The equation \eqref{matrixPowersSumAdjacencyConclusion} implies that $\overline{\mathcal{O}}$ is an adjacency matrix of a graph defined by the paths of lengths: $0,1,\ldots, p$ in $\mathbb{G}$}. Furthermore, like it is mentioned in Section \ref{liftedRepresentationSection}, the lifted equation \eqref{liftedClassicalOutput} has a graph theoretic interpretation. Namely, the local output $\mathbf{y}_{i}(k)$ depends on $\mathbf{x}_{j}(k-p)$ if there is a path of length $p$ from the vertex $i$ to the vertex $j$ in $\mathbb{G}$. \textit{That is, by increasing $p$ in \eqref{liftedClassicalOutput}, we are actually reaching the states of local subsystems (vertices) that are further away from the local subsystem $\mathcal{S}_{i}$ (vertex $i$)}. $\hfill \square$
\label{remarkGraphTheory} 
\end{rem}
Similarly, we can assign to $\mathcal{G}_{p}$ a binary matrix describing its block sparsity pattern. Let such a matrix be denoted by $\overline{\mathcal{G}}$. It can be easily shown that the sparsity structure of $\overline{\mathcal{G}}$ is determined by the powers of $\underline{A}$. Next, we analyze the sparsity pattern of $\mathcal{W}$. Analogously to the matrix $\overline{\mathcal{O}}$ that describes the sparsity pattern of $\mathcal{O}_{p}$, we can define a matrix $\overline{\mathcal{W}}$ describing the block sparsity pattern of $\mathcal{W}$.
\begin{prop}
The sparsity pattern of $\overline{\mathcal{W}}$ is given by
\begin{align}
\overline{\mathcal{W}}=T\left( \sum_{i=0}^{p}\left(\overline{A}^{T}\right)^{i}\overline{A}^{i}\right)
\label{proof1}
\end{align}
\end{prop}
\textit{Proof}. The finite-time observability Gramian is defined by summing up the following terms (see the equation \eqref{obsGrammian}):
\begin{align}
\left(\underline{A}^{T} \right)^{i}\underline{C}^{T}\underline{C} \underline{A}^{i}
\label{termDefinitionGramian}
\end{align}
The sparsity pattern of \eqref{termDefinitionGramian} is determined by the sparsity pattern of $\left(\overline{A}^{T} \right)^{i}\overline{C}^{T}\overline{C} \overline{A}^{i}$ or equivalently, by the sparsity pattern of $\left(\overline{A}^{T} \right)^{i}\overline{A}^{i}$. This further implies that the block sparsity pattern of $\mathcal{W}$ can be described by the non-zero elements of:
\begin{align}
\sum_{i=0}^{p}\left(\overline{A}^{T}\right)^{i}\overline{A}^{i}
\label{proof2}
\end{align}
By applying the operator $T(\cdot)$ to \eqref{proof2}, we obtain \eqref{proof1}. $\hfill \square$
\par
It should be noted that the matrix $\overline{A}^{T}$ is an adjacency matrix of a graph obtained by transposing the graph $\mathbb{G}$ (transpose of the graph $\mathbb{G}$ is obtained by reversing the edges of $\mathbb{G}$). If the graph $\mathbb{G}$ is undirected, then the interconnection matrix $\overline{A}$ is symmetric. Consequently, from \eqref{proof1} we can conclude that in the case of the undirected graph $\mathbb{G}$, the structure of $\overline{\mathcal{W}}$ is determined by the paths of lengths $2,4,\ldots,2p$ in $\mathbb{G}$.
\par
Consider the first sparsification operator of the Newton-Schultz iteration and let the sparsity pattern of the approximate inverse $X$ be described by the matrix $P$. Let us assume that the sparsity pattern of $P$ is equal to the sparsity pattern of a matrix obtained by summing up the first $s$ powers of $\mathcal{W}$ (see Section \ref{SparsityPatternSelection}). This implies that the sparsity pattern of $X$ is described by the sparsity pattern of
\begin{align} \overline{X}=T\left(I+\overline{\mathcal{W}}+\ldots+\overline{\mathcal{W}}^{s} \right)
\label{interconnectionX}
\end{align}
\textit{That is, $\overline{X}$ is an adjacency matrix of a graph defined by the paths of lengths $1,\ldots, s$ in the graph whose adjacency matrix is $\overline{\mathcal{W}}$, see Remark \ref{remarkGraphTheory}}. 
Similarly, if the Newton-Schultz iteration is used without the sparsification operators, then the sparsity pattern of $\overline{X}$ is determined by the sum of powers of $\overline{\mathcal{W}}$. Namely, by propagating \eqref{NewtonSchultz} it can be shown that $X_{k}$ can be expressed as a sum of powers of $\mathcal{W}$. 
\par
Consider the distributed estimator \eqref{approximateInverseSolution}. Let the block sparsity patterns of $L$ and $Q$ be described by binary matrices $\overline{L}$ and $\overline{Q}$, respectively. \textit{Graphs whose adjacency matrices are $\overline{L}$ and $\overline{Q}$ describe the communication links between the local subsystems of the distributed estimator \eqref{approximateInverseSolution}.} From \eqref{approximateInverseSolution} we have that the sparsity pattern of $\overline{L}$ is determined by the product of $\overline{X}$ and $\overline{\mathcal{O}}^{T}$. Similarly, the structure of $\overline{Q}$ is determined by the product of $\overline{X}$, $\overline{\mathcal{O}}^{T}$ and $\overline{\mathcal{G}}$.
From \eqref{proof1} and \eqref{interconnectionX} we see that the sparsity pattern of $\overline{X}$ is determined by the sparsity pattern of $\overline{A}$. On the other hand, from \eqref{matrixPowersSumAdjacencyConclusion} we also conclude that the structure of $\overline{\mathcal{O}}$ is determined by $\overline{A}$ (using the same principle that was used to derive \eqref{matrixPowersSumAdjacencyConclusion}, it can be shown that the structure of $\overline{\mathcal{G}}$ is determined by $\overline{A}$). All this shows that the structure of $\overline{L}$ and $\overline{Q}$ is determined by the structure of $\overline{A}$. If the interconnection structure of $\mathcal{S}$ is described by undirected graph $\mathbb{G}$, then the structure of $\overline{L}$ and $\overline{Q}$ is determined by the paths in $\mathbb{G}$.
\section{Numerical simulations}
\label{numericalSumulations}
\begin{figure}[H]
\centering 
\includegraphics[scale=0.27,trim=0mm 0mm 0mm 0mm,clip=true]{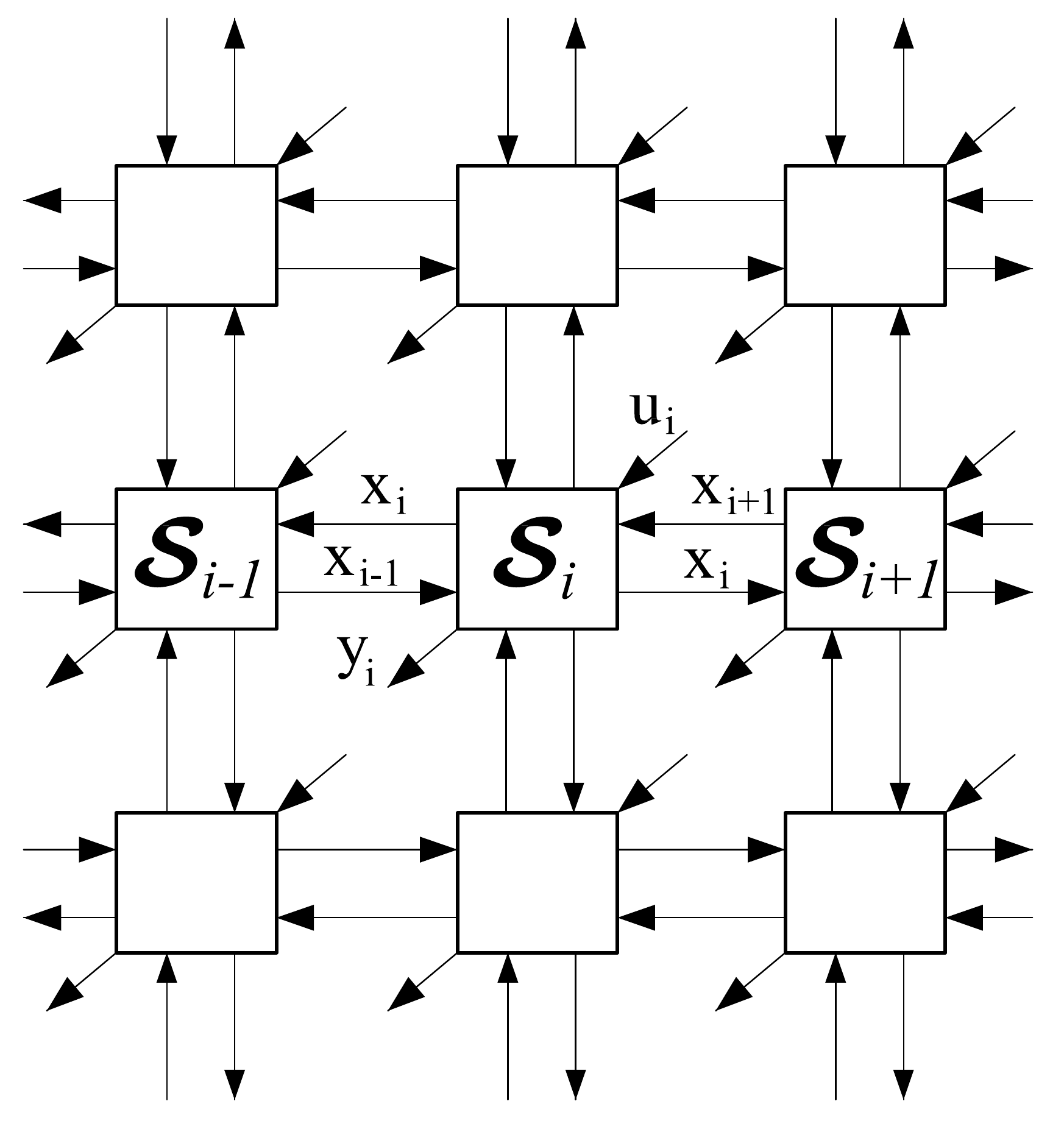}
\caption{Interconnection pattern of local subsystems.}
\label{interconnectionPattern}
\end{figure}

\begin{figure}[H]
  \centering
    \subfloat[]{\includegraphics[scale=0.18,trim=0mm 0mm 0mm 0mm ,clip=true]{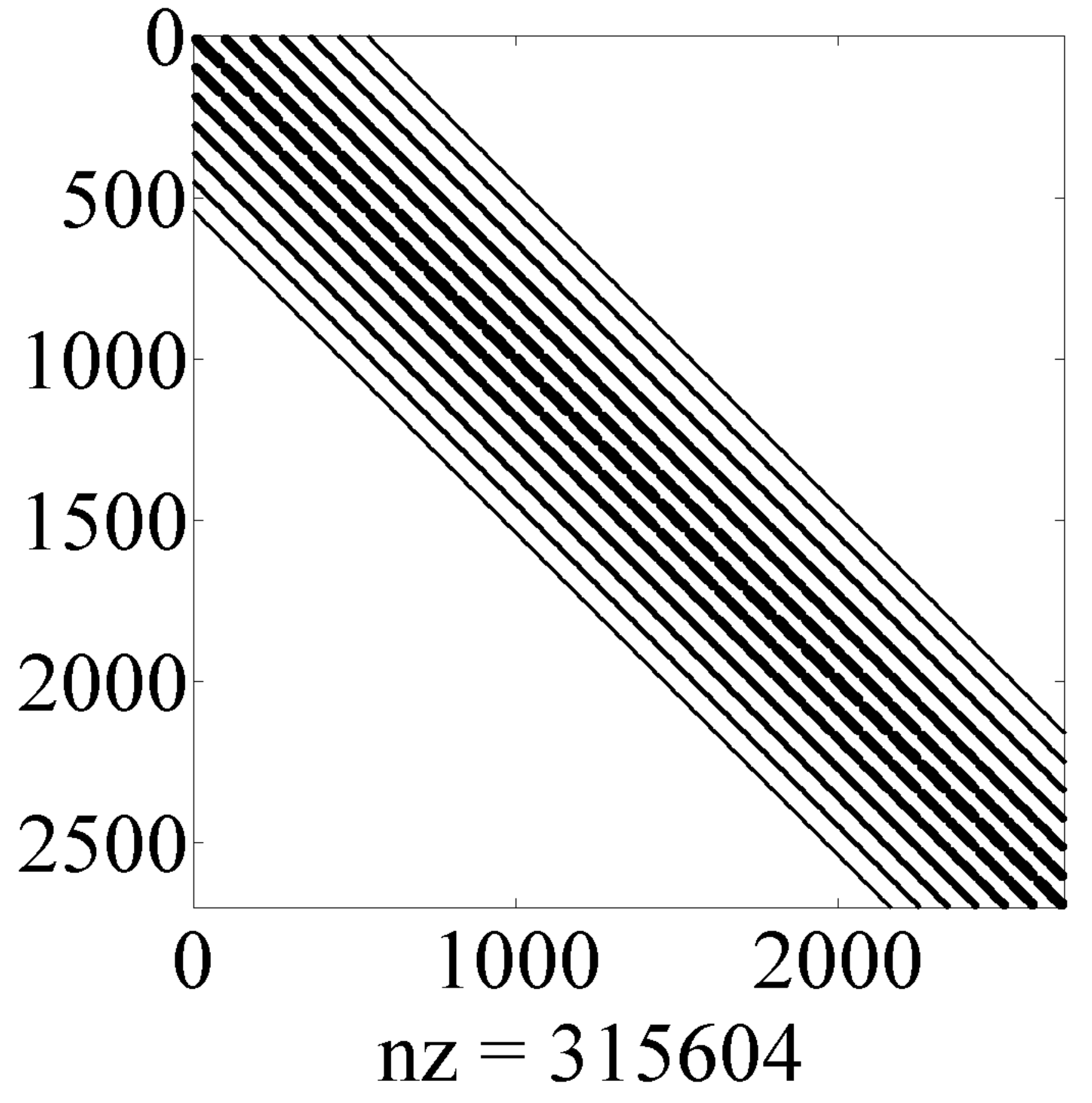}}
   \subfloat[]{\includegraphics[scale=0.17,trim=0mm 0mm 0mm 0mm ,clip=true]{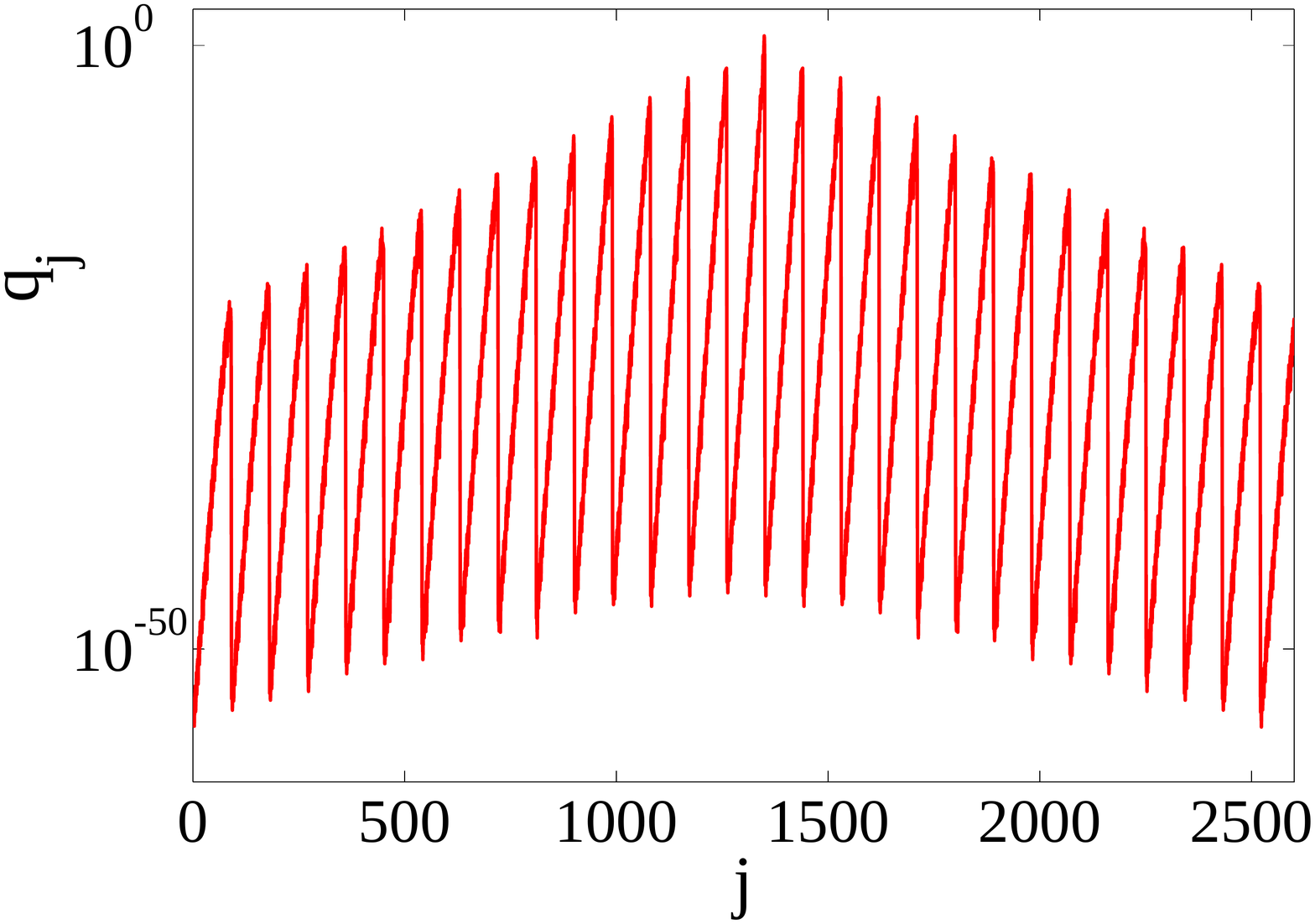}}
  \caption{ (a) Sparsity pattern of $\mathcal{W}_{r}$, "nz" denotes the number of non-zero elements; (b) The absolute values of elements of an arbitrary row of $\mathcal{W}_{r}^{-1}$.}
\label{fig:gramMulti}
\end{figure}

\begin{figure}[H]
  \centering
 \subfloat[]{\includegraphics[scale=0.18,trim=0mm 0mm 0mm 0mm ,clip=true]{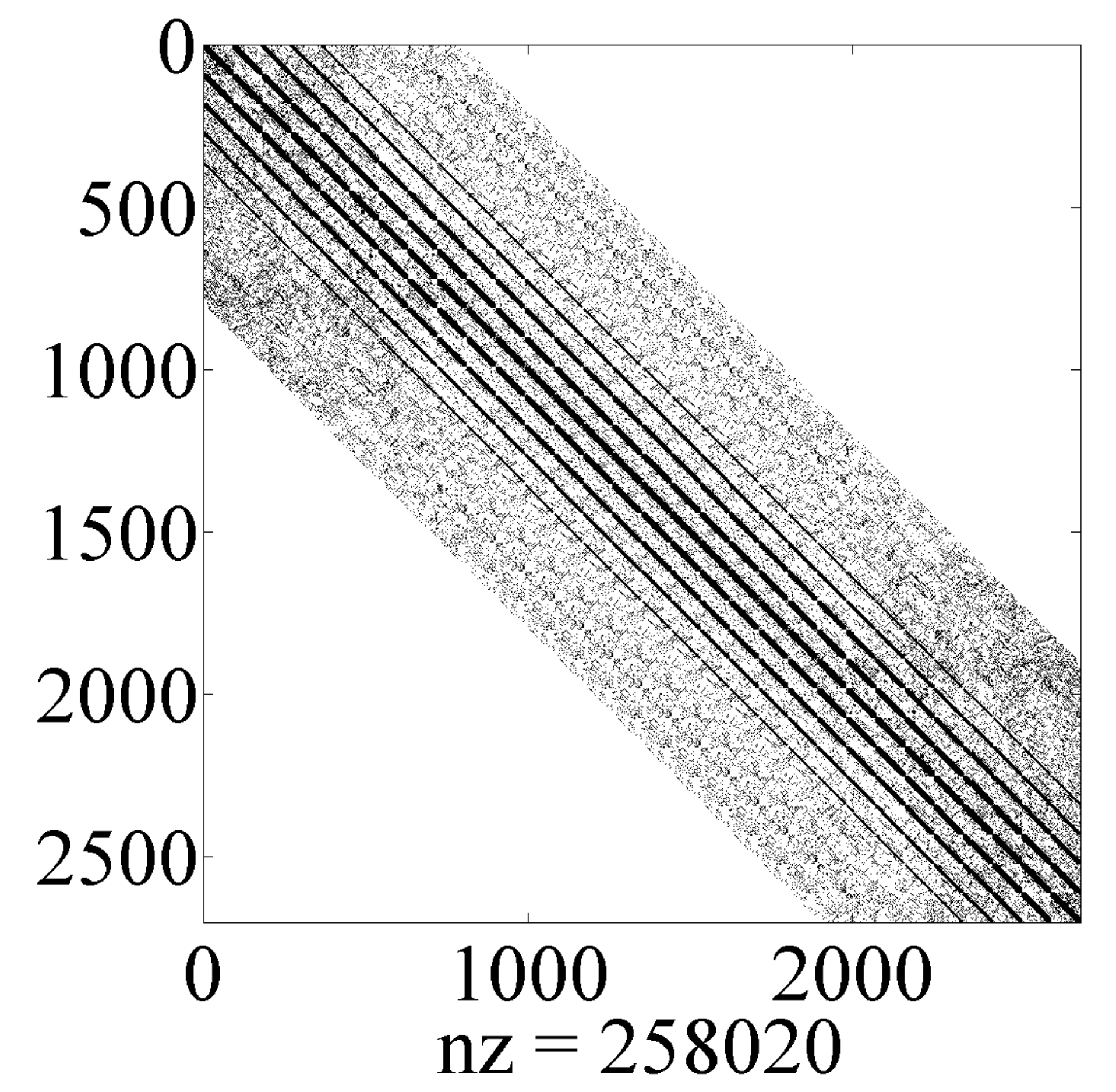}} \;\;\;
\subfloat[]{\includegraphics[scale=0.18,trim=0mm 0mm 0mm 0mm ,clip=true]{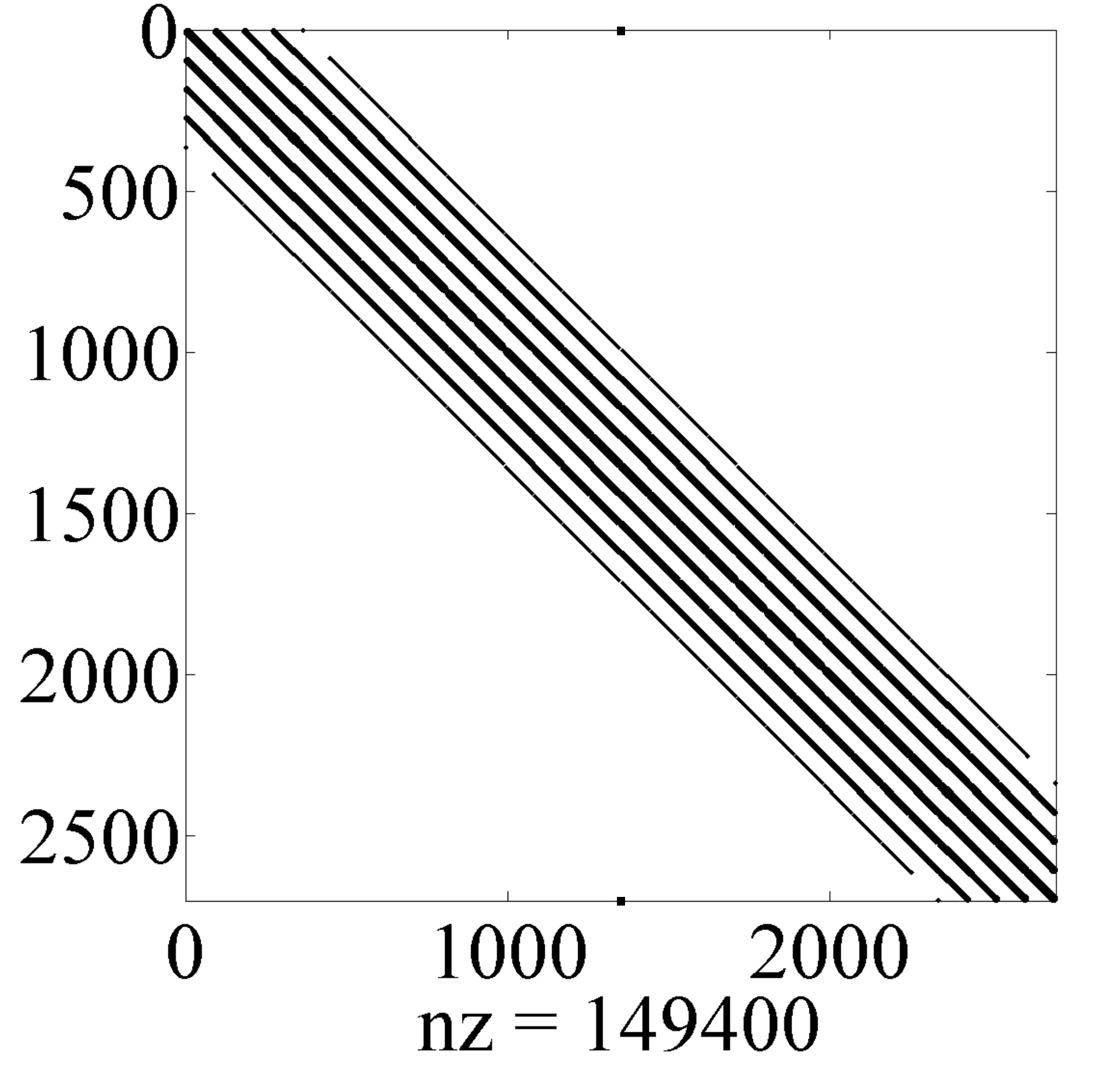}} 
  \caption{Sparsity patterns of approximate inverses of $\mathcal{W}_{r}$: (a) $\beta=800$, $\phi=0.00001$, $e=0.0099$ ; (b) $\beta=800$, $\phi=0.00005$, $e=0.0103$.}
\label{fig:offDiagonalDecayGr}
\end{figure}

\begin{figure}[H]
\centering
\subfloat[]{\includegraphics[scale=0.22,trim=0mm 0mm 0mm 0mm ,clip=true]{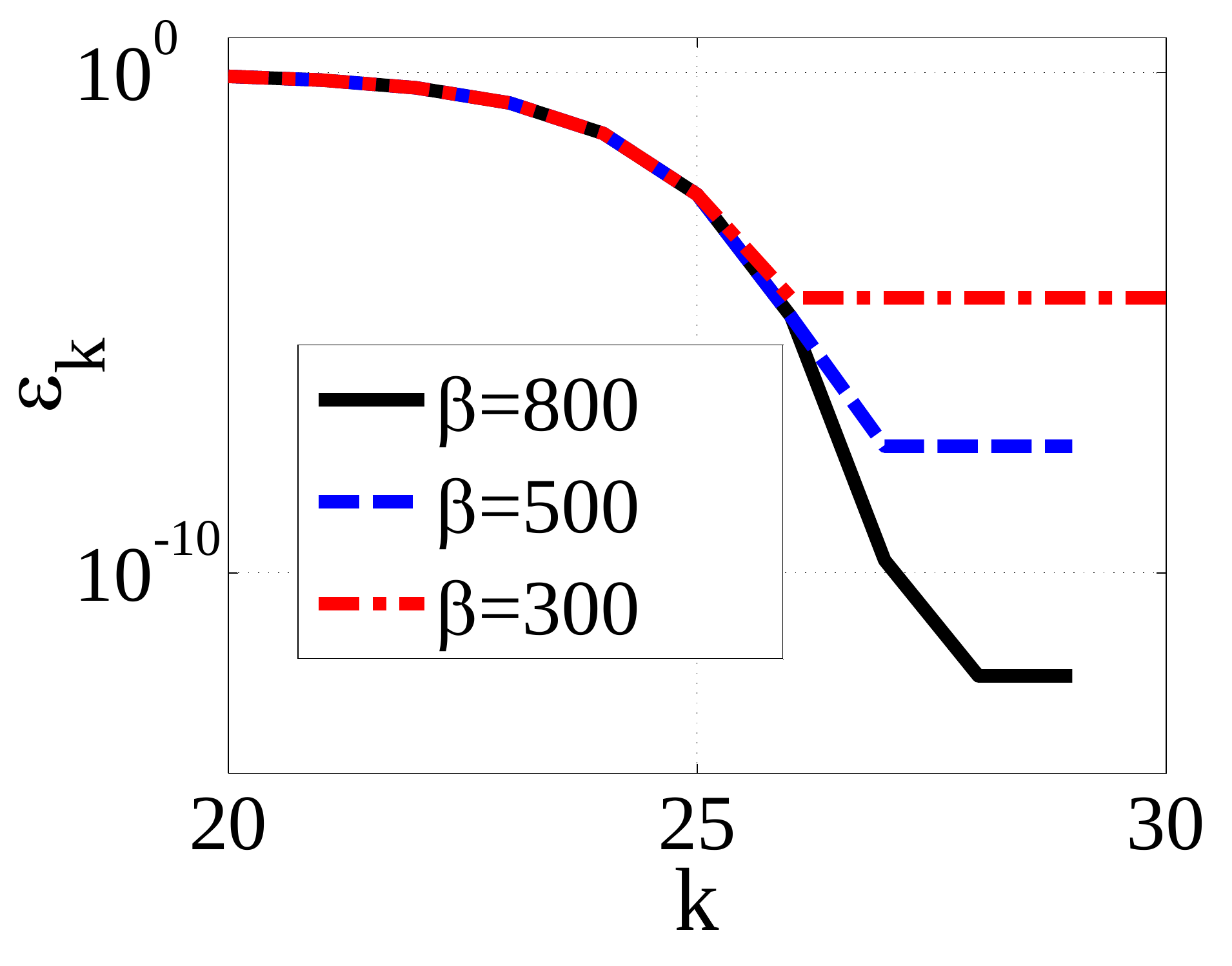}}
\subfloat[]{\includegraphics[scale=0.18,trim=0mm 0mm 0mm 0mm ,clip=true]{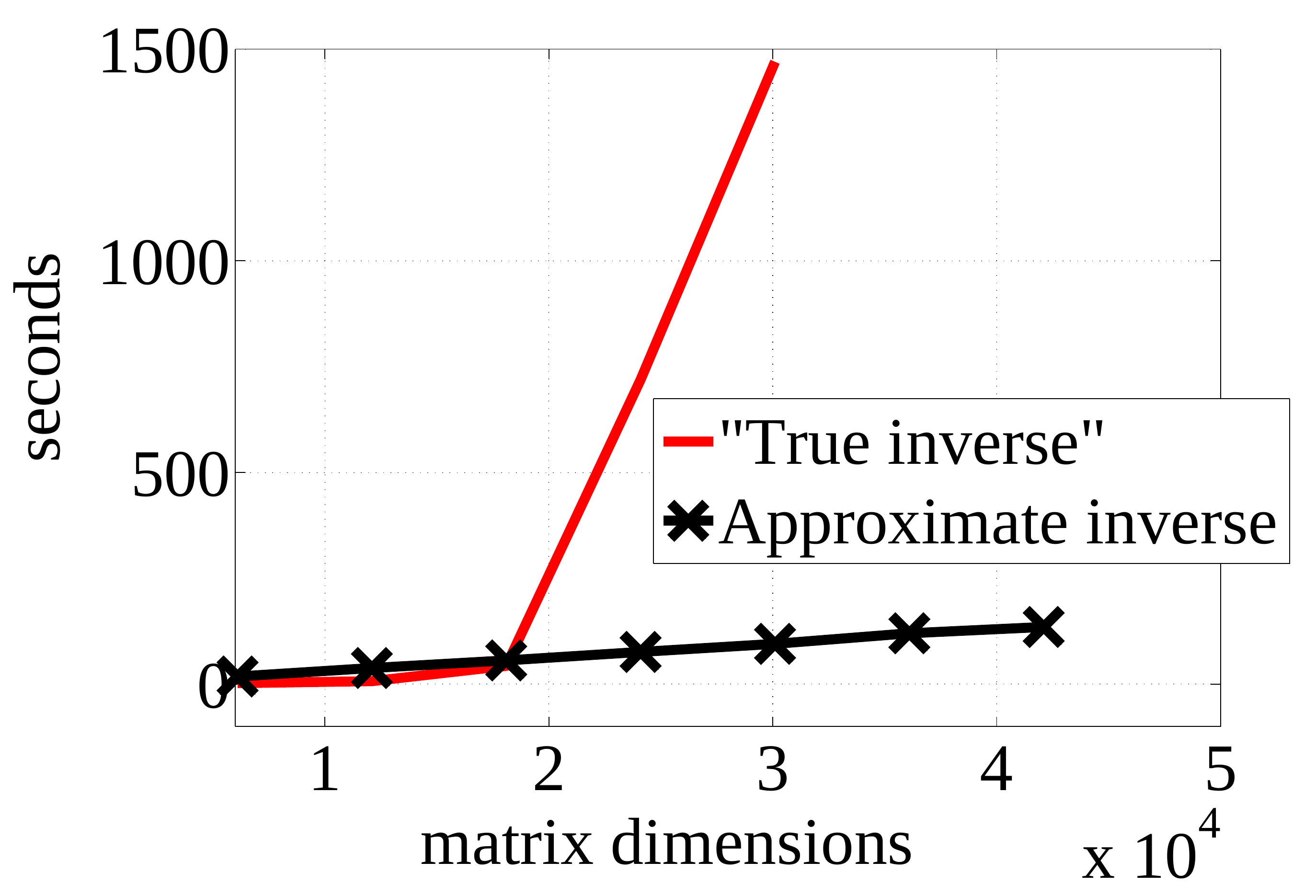}} 
\caption{(a) The convergence of the Newton iteration for several values of $\beta$. (b) Computational complexity of the Newton-Schultz iteration compared to "built-in" MATLAB inversion function.}
\label{fig:newtonFinal}
\end{figure}

We are considering the problem of approximating a regularized, finite-time, observability Gramian of a state-space model obtained by discretizing the 3D heat equation. The state-space model used in this paper is derived in Chapter 2.3 of \cite{haberThesis}. The model consists of the interconnection of 900 local subsystems. The interconnection pattern of local subsystems is shown in Fig. \ref{interconnectionPattern}. The local state order of each subsystem is $n=3$. Each local subsystem has one output and one input. In total, the global system has 2700 states, 900 inputs and 900 outputs.
 We form the matrix $\mathcal{O}_{p}$ for $p=4$. Figure \ref{fig:gramMulti}(a) shows the sparsity pattern of the regularized, finite-time observability Gramian $\mathcal{W}_{r}=0.001 I+\mathcal{O}_{4}^{T}\mathcal{O}_{4}$. The condition number of $\mathcal{W}_{r}$ is $\kappa=3.78\times 10^{3}$. Because the dimensions of $\mathcal{W}_{r}$ are not extremely large, we are able to compute $\mathcal{W}_{r}^{-1}$. The "true" inverse $\mathcal{W}_{r}^{-1}$ helps us to quantify the accuracy of the proposed approximation framework. Figure \ref{fig:gramMulti}(b) shows absolute values of elements of an arbitrary row of $\mathcal{W}_{r}^{-1}$. 
Figure \ref{fig:gramMulti}(b) shows that $\mathcal{W}_{r}^{-1}$ is an off-diagonally decaying matrix. Furthermore, Fig. \ref{fig:gramMulti}(b) shows that the rate of the off-diagonal decay of $\mathcal{W}_{r}^{-1}$ is fast, which is expected because the matrix $\mathcal{W}_{r}$ is relatively well-conditioned. All this indicates that there exists a sparse approximate inverse of $\mathcal{W}_{r}$. We use the sparsified Newton-Schultz iteration to compute $X$ (we combine both sparsification operators). The final approximation error is quantified by: $e=\left\|\mathcal{W}_{r}^{-1}-X\right\|_{2}$. The pattern matrix $P$ is a banded matrix, with the bandwidth equal to $\beta$. For example, the bandwidth of the matrix that is illustrated in Fig. \ref{fig:gramMulti}(a) is 550. Figure \ref{fig:offDiagonalDecayGr} shows the sparsity patterns of approximate inverses calculated for $\beta=800$ and for two values of $\phi$ ($\phi$ is the dropping parameter, see equation \eqref{eq:operator1}).
By comparing Fig. \ref{fig:gramMulti}(a) and Fig. \ref{fig:offDiagonalDecayGr} we see that the approximate inverses have sparsity patterns that are similar to the sparsity pattern of $\mathcal{W}_{r}$. Furthermore, from Fig. \ref{fig:offDiagonalDecayGr} we see that by increasing $\phi$ we can additionally sparsify the approximate inverse. However, as we increase $\phi$, the final approximation error increases.
\\
Figure \ref{fig:newtonFinal}(a) shows how the errors introduced by the sparsification operator $H_{1}$ influence the convergence of the Newton-Schultz iteration. The approximation error is calculated in each iteration as follows: $\epsilon_{k}=\left\|I-\mathcal{W}_{r}H_{1}\left(Z_{k}\right) \right\|_{2}$. We see that as $\beta$ decreases, the final value of the approximation error increases and vice-versa. Furthermore, we see that $\beta$ does not significantly decrease the convergence rate of the Newton-Schultz iteration. We have observed that for relatively small $\beta$, the Newton-Schultz iteration starts to diverge. 
Finally, in Fig. \ref{fig:newtonFinal}(b) we compare the computational complexity of the proposed approximation algorithm with the complexity of the "built-in" MATLAB inversion function. We vary the dimensions of the observability Gramian and we measure the time necessary to invert it. The MATLAB inversion function is exploiting the sparsity of the problem, and its complexity scales at least quadratically with the size of $\mathcal{W}_{r}$. On the other hand, the computational complexity of the proposed approximation framework is linear. For brevity, we omit a figure showing the memory complexity of the proposed method. Memory complexity of our method is linear, in contrast to quadratic memory complexity of MATLAB inversion algorithm.
\section{Conclusion and future work}
\label{conclusionSection}
In this paper we showed that the inverses of the finite-time observability Gramians and impulse response matrices can be approximated by sparse matrices with linear computational and memory complexity. This result opens the door to the development of novel approach to estimation and control of large-scale systems. The novel estimators (controllers) compute local estimates (control actions) simply as a linear combination of local data. The size of the local data set depends on the condition number of the finite-time Gramians. The numerical results confirmed the effectiveness of the proposed approximation framework. An important goal of future work should be to analyze the errors introduced by the sparsification operators. Furthermore, the future work should address the following question. \textit{Can the established approximation methodology be used to approximate the solution of the Riccati equation by a sparse structured matrix?} The affirmative answer to this question would mean that it is possible to efficiently compute distributed LQG controllers for large-scale dynamical networks.
\bibliographystyle{unsrt}
\bibliography{bibl}
\end{document}